\documentclass[english,aps,prl,amsmath,twocolumn,preprintnumber,superscriptaddress,showpacs,notitlepage]{revtex4-2}
\usepackage[T1]{fontenc}
\usepackage[latin9]{inputenc}
\DeclareMathAlphabet\mathbfcal{OMS}{cmsy}{b}{n}

\makeatletter

\usepackage{mathptmx,newtxtext,newtxmath,xspace}
\usepackage{amsbsy,bm,bbold}
\usepackage{graphicx,color,xcolor,epsfig,rotate}
\usepackage{fancyhdr}
\usepackage[colorlinks=true, 
            linkcolor=blue, 
            urlcolor=blue,
            citecolor=blue]{hyperref}
\usepackage{soul}
\usepackage{cancel}
\newcommand{\iu}{{i\mkern1mu}}

\makeatother

\usepackage{babel}
\begin{document}
\title{\texorpdfstring{CP$^2$}{CP2} Skyrmions and Skyrmion Crystals in Realistic Quantum Magnets}
\author{Hao~Zhang}
\affiliation{Department of Physics and Astronomy, The University of Tennessee,
Knoxville, Tennessee 37996, USA}
\affiliation{Materials Science and Technology Division, Oak Ridge National Laboratory, Oak Ridge, Tennessee 37831, USA}
\author{Zhentao~Wang}
\affiliation{Department of Physics and Astronomy, The University of Tennessee, Knoxville, Tennessee 37996, USA}
\affiliation{School of Physics and Astronomy, University of Minnesota, Minneapolis, Minnesota 55455, USA}
\author{David~Dahlbom}
\affiliation{Department of Physics and Astronomy, The University of Tennessee,
Knoxville, Tennessee 37996, USA}
\author{Kipton~Barros}
\affiliation{Theoretical Division and CNLS, Los Alamos National Laboratory, Los Alamos, New Mexico 87545, USA}
\author{Cristian~D.~Batista}
\affiliation{Department of Physics and Astronomy, The University of Tennessee,
Knoxville, Tennessee 37996, USA}
\affiliation{Quantum Condensed Matter Division and Shull-Wollan Center, Oak Ridge
National Laboratory, Oak Ridge, Tennessee 37831, USA}
\date{\today}
\begin{abstract}
Magnetic skyrmions are nanoscale topological textures that have been recently observed in different families of quantum magnets. These textures are known as CP$^1$ skyrmions because the target manifold of the magnetization field is the 2D sphere $S^2 \cong$ CP$^{1}$. 
Here we report the emergence of magnetic CP$^2$ skyrmions in realistic spin-$1$ models. Unlike  CP$^1$  skyrmions, CP$^2$ skyrmions can also arise as metastable textures of quantum paramagnets, opening a new road to discover emergent topological solitons in non-magnetic materials. The quantum phase diagram of the spin-$1$ models also  includes  magnetic field-induced CP$^2$ skyrmion crystals that can be detected with regular momentum- (diffraction) and real-space (Lorentz transmission electron microscopy) experimental techniques. 
\end{abstract}
\maketitle

\section{Introduction}

Lord Kelvin's vision of the atom as a vortex in ether~\cite{Thomson67}
inspired Skyrme~\cite{Skyrme61,Skyrme88}  to explain the origin of nucleons as emergent topologically non-trivial configurations of  a  pion field described by a 3+1 dimensional O(4) non-linear $\sigma$-model. In the modern language, these ``skyrmions'' are examples of topological solitons and Skyrme's model has become the prototype of a classical theory that supports these solutions. Besides its important role in  high-energy physics and cosmology, Skyrme's model also led to important developments in other areas of physics. For instance, the baby Skyrme model~\cite{Bogolubskaya89,Bogolyubskaya89b,Leese90} (planar reduction of the non-linear $\sigma$-model), which is as an extension of the Heisenberg model~\cite{Bogolubskaya89,Bogolyubskaya89b,Polyakov75}, has baby skyrmion solutions in the presence of a chiral symmetry breaking Dzyaloshinskii-Moriya (DM) interaction~\cite{Dzyaloshinsky1958,Moriya60b,Moriya1960,Bogdanov94}.

Periodic arrays of  magnetic skyrmions and single skyrmion metastable states were originally observed  in chiral magnets, such as MnSi, Fe$_{1-x}$Co$_x$Si, FeGe and Cu$_2$OSeO$_3$~\citep{Muhlbauer2009, Yu2010a,Yu2011,Seki2012,Adams2012}. This discovery sparked the interest of the  community at large and spawned efforts in multiple directions. Identifying realistic conditions for the emergence of novel magnetic skyrmions  is one of the main goals of modern condensed matter physics. Novel mechanisms are usually  accompanied by new properties. For instance, while skyrmions of chiral magnets have a fixed vector chirality,  this is still a degree of freedom in centrosymmetric materials, such as BaFe$_{1-x-0.05}$Sc$_x$Mg$_{0.05}$O$_{19}$, La$_{2-2x}$Sr$_{1+2x}$Mn$_2$O$_7$, Gd$_2$PdSi$_3$ and Gd$_3$Ru$_4$Al$_{12}$~\cite{Yu2012_BFSMO,Yu2014_biskyrmion,Mallik1998_paramana,Saha1999,Kurumaji2019,Chandragiri_2016,Hirschberger2019}, where skyrmions arise from frustration, i.e., from competing exchange  or dipolar interactions~\citep{Okubo12,Leonov2015,Lin2016_skyrmion,Hayami16,Batista16,Wang2020_RKKY,Hayami2021_review}.

The target manifold of the above-mentioned planar baby skyrmions is $S^2 \cong$ CP$^{1}$, i.e., the usual 2D sphere, corresponding to normalized dipoles. More generally, one may consider the complex projective space CP$^{N-1}$ that represents the normalized $N$-component complex vectors, up to an irrelevant complex phase. The topologically distinct, smooth mappings from the base manifold $S^2$ (2D sphere $\cong$ compactified plane)  to the target manifold CP$^{N-1}$ can be labeled by the integers: 
$\Pi_2(\mathrm{CP}^{N-1})= \mathbb{Z}$. This homotopy group suggests generalizations of the planar Skyrme's model to $N > 2$, such as the CP$^2$ non-linear $\sigma$-model~\cite{Golo78,Adda78,Din80} and in the Faddeev-Skyrme type model~\cite{Ferreira10,Amari15}. In recent work, Akagi et al.\ considered the SU(3) version  of the Heisenberg model with a DM interaction, whose continuum limit becomes a gauged CP$^2$ nonlinear $\sigma$-model with a background uniform gauge field~\cite{Akagi21}. An attractive aspect of this model is that it admits analytical solution by the application of techniques developed for the gauged non-linear $\sigma$-model. However, it may be challenging to find materials described by this model because SU(3) can only be an accidental symmetry of spin-spin interactions of real quantum magnets.

The main purpose of this work is to demonstrate that exotic CP$^2$ skyrmions readily emerge in simple and realistic spin-$1$ ($N=3$) models and their higher-spin generalizations. Remarkably,  isolated CP$^2$ skyrmions can either be  metastable  states of a quantum paramagnet (QPM) or of a fully polarized (FP) ferromagnet. Unlike the ``usual'' CP$^{1}$ magnetic skyrmions, the dipolar field of the metastable CP$^2$ skyrmions of quantum paramagnets vanishes away from the skyrmion core. Moreover, the application of an external magnetic field to the QPM induces  stable triangular crystals of  CP$^2$ skyrmions  in the field interval that separates the QPM from the FP state.

\section{Model}

To illustrate the basic ideas we consider a minimal spin-$1$ model defined on the triangular lattice (TL):
\begin{equation}
\!\!\! {\hat {\mathcal H}}= \sum_{\langle i,j \rangle} J_{ij} \left( {\hat S}^x_i {\hat S}^x_j + {\hat S}^y_i {\hat S}^y_j + \Delta {\hat S}^z_i {\hat S}^z_j \right) - h \sum_{i}  {\hat S}^z_i + D \sum_i \left({\hat S}^z_i \right)^2.
\label{hamil}
\end{equation}
The first term includes an easy-axis ferromagnetic (FM) nearest-neighbor exchange interaction $J_1<0$ and a second-nearest-neighbor antiferromagnetic (AFM) exchange $J_2>0$. For simplicity, we assume that the exchange anisotropy, defined by the dimensionless parameter $\Delta>1$, is the same for both  interactions. The second and third terms represent the Zeeman coupling to an external field and an easy-plane  single-ion anisotropy ($D>0$), respectively. $\hat{\mathcal{H}}$ is invariant under the space-group of the TL and  the U(1) group of  global spin rotations along the field-axis. We will adopt $|J_1|$ as the unit of energy (i.e. $J_1=-1$).

The first step is to take the classical limit of $\hat{\mathcal{H}}$, where multiple approaches are possible~\cite{Gnutzmann98, Hao21}.
The traditional classical limit is based on SU(2) coherent states, which retains only the spin dipole expectation value, and produces the Landau-Lifshitz spin dynamics. This approach is adequate for modeling systems with weak single-ion anisotropy $D \ll |J_1|$. To model systems in the regime  $D \gtrsim |J_1|$, and correctly capture low-energy excitations, it is important to retain more information about spin fluctuations in the $(N=3)$-dimensional local Hilbert space. Specifically, our classical limit will assume that the many-body quantum state is a 
direct product of SU(3) coherent states~\cite{Papanicolaou1988,Gnutzmann98,Batista04,Zapf06,Lauchli2006,Muniz14,Hao21}:
\begin{equation}
|\bm{Z} \rangle=\otimes_{j}|\bm{Z}_j \rangle
\;\; {\rm with} \;\;
|\bm{Z}_{j} \rangle=\sum_a Z^{a}_j \left|x^{a}\right\rangle_{j},
\label{eq:z}
\end{equation}
where $\bm{Z}_j=\left(Z^{1}_j, Z^{2}_j, Z^{3}_j \right)^{\mathrm{T}}$ is a complex vector of unit length and $\{ \left|x^{1}\right\rangle_{j}, \left|x^{2}\right\rangle_{j}, \left|x^{3}\right\rangle_{j} \}$ is an orthonormal basis for the local Hilbert state on site $j$. 

Local physical operators are represented by Hermitian matrices that act on SU(3) coherent states. The space of $3\times 3$ traceless, Hermitian matrices comprises the fundamental representation of the $\mathfrak{su}(3)$ Lie algebra. A basis $\hat T^\mu$ ($\mu=1,\dots,8$) for this space is characterized by the commutation relations,
\begin{equation}
\left[{\hat T}_{j}^{\eta}, {\hat T}^{\mu}_j \right]=i f_{\eta \mu \nu} {\hat T}_{j}^{\nu},\label{eq:bracket}
\end{equation}
where we are using the convention of summation over repeated Greek indices. We may additionally impose an orthonormality condition
\begin{equation}
\operatorname{Tr}\left({\hat T}_j^{\alpha} {\hat T}_j^{\beta}\right)=2 \delta_{\alpha \beta}.\label{eq:ortho}
\end{equation}
It is conventional to define the structure constants as 
$
f_{\eta \mu \nu}=-\frac{i}{2} \operatorname{Tr}\left({\lambda}_{\eta}\left[{\lambda}_{\mu}, {\lambda}_{\nu}\right]\right),
$
where ${\lambda}_\mu$ are the Gell-Mann matrices.

The spin dipole operators $\hat{\boldsymbol S}_j = (\hat{S}^x_j,\hat{S}^y_j,\hat{S}^z_j)^T$ acting on site $j$ are generators for a spin-$1$ representation of SU(2). It is possible to formulate generators of SU(3) as polynomials of these spin operators,
\begin{eqnarray}
\left(\begin{array}{l}
{\hat T}_j^{7} \\
{\hat T}_j^{5} \\
{\hat T}_j^{2}
\end{array}\right)=- \left(\begin{array}{c}
{\hat S}_j^{x} \\
{\hat S}_j^{y} \\
{\hat S}_j^{z}
\end{array}\right), \left(\begin{array}{l}
{\hat T}_j^{3} \\
{\hat T}_j^{8} \\
{\hat T}_j^{1} \\
{\hat T}_j^{4} \\
{\hat T}_j^{6}
\end{array}\right)=\left(\begin{array}{c}
-\left({\hat S}_j^{x}\right)^{2}+\left({\hat S}_j^{y}\right)^{2} \\
\frac{1}{\sqrt{3}}\left[ 3\left({\hat S}_j^{z}\right)^{2} - {\hat {\boldsymbol S}}_j^2 \right] \\
 {\hat S}_j^{x} {\hat S}_j^{y}+{\hat S}_j^{y} {\hat S}_j^{x} \\
-{\hat S}_j^{z} {\hat S}_j^{x}-{\hat S}_j^{x} {\hat S}_j^{z} \\
 {\hat S}_j^{y} {\hat S}_j^{z}+{\hat S}_j^{z} {\hat S}_j^{y},
\end{array}\right),
\label{eq:gens}
\end{eqnarray}
where $T_j^{7,5,2}$ are the dipolar components of the spin-1 degree of freedom, while  the other five generators are the quadrupolar components. 
Here we have adopted the notation and conventions of Ref.~\cite{Akagi21} to make closer contact with the literature on high-energy physics~\footnote{Our definitions for $\hat{S}^x$ and $\hat{S}^z$ differ from these two in Ref~\citep{Akagi21} by a minus sign.}.

Let $| 1\rangle_j$, $|0 \rangle_j$, and $|\bar{1} \rangle_j$ denote the normalized eigenstates of ${\hat S}^z_j$, with eigenvalues, 1, 0 and -1, respectively. In the Cartesian basis,
\begin{equation}
\left|x^{1}\right\rangle_j =
\frac{i \left[ |1\rangle_j -|\bar{1} \rangle_j \right]}{\sqrt{2}}, \left|x^{2}\right\rangle_j= \frac{
\left [ |1\rangle_j +|\bar{1} \rangle_j \right]}{\sqrt{2}},
\left|x^{3}\right\rangle_j = -i|0\rangle_j,
\end{equation}
the SU(3) generators given in Eq.~\eqref{eq:gens} are  the Gell-Mann matrices:
\begin{equation}
\langle x^a_j | {\hat T}_j^{\mu} | x^b_j \rangle =\left({\lambda}_{\mu}\right)_{a b} \quad \mu=1,2, \cdots, 8.
\end{equation}

The orbit of coherent states $|\bm{Z}_j \rangle$ is obtained by applying SU(3) transformations to the highest weight state $|1\rangle_j$~\cite{Gnutzmann98}: $|\bm{Z}_j \rangle = {\hat U}_j |1 \rangle_j$.
Since the global phase is a gauge degree of freedom, the orbit of physical SU(3) coherent states is $S^{5} / S^{1} \cong \mathrm{CP}^{2}$. 
The ``SU(3) classical limit'' of the spin Hamiltonian \eqref{hamil} is obtained by replacing the Hamiltonian operator ${\hat {\mathcal H}}$ with its expectation value
\begin{equation}
{\cal H} \equiv \langle \bm{Z}| {\hat {\mathcal H}} | \bm{Z}\rangle,
\end{equation}
after rewriting ${\hat {\mathcal H}} $ in terms of SU(3) spin components,
\begin{equation}
{\hat {\mathcal H}}= \sum_{\langle i,j \rangle} {I}^{\mu }_{ij}  {\hat T}^{\mu}_i {\hat T}^{\mu}_j  - \sum_i B^{\mu} {\hat T}^{\mu}_i,
\label{hamil2}
\end{equation}
where ${I}^{\mu }_{ij} = J_{ij}  (\Delta \delta_{\mu,2}+\delta_{\mu,5}+\delta_{\mu,7})$ and $B^{\mu} =(-h \delta_{\mu,2}-D \delta_{\mu,8}/\sqrt{3})$.
Because of the direct product form of  Eq.~\eqref{eq:z}, ${\cal H} $ can be expressed as a function  of the ``color field''
\begin{equation}
n^{\mu}_j \equiv \langle \bm{Z}_j| {\hat T}_{j}^{\mu} |\bm{Z}_j\rangle =\left({\lambda}_{\mu}\right)_{a b} \bar{Z}^{a}_j Z^{b}_j,
\label{eq:cfield}
\end{equation}
which satisfies the constraints
\begin{equation}
n^{\mu} n^{\mu}=\frac{4}{3}, \quad n^{\mu}=\frac{3}{2} d_{\mu \nu \eta} n^{\nu} n^{\eta},
\label{eq:const}
\end{equation}
where $d_{\mu \nu \eta}=\frac{1}{4} \operatorname{Tr}\left(\lambda_{\mu}\left\{\lambda_{\nu}, \lambda_{\eta}\right\}\right)$. 
This in turn leads to the Casimir identity:
$
d_{m p q} n^{m} n^{p} n^{q}= \frac{8}{9}.
$
In terms of this color field, we can express
\begin{equation}
{\mathcal H}= \sum_{\langle i,j \rangle} {I}^{\mu }_{ij}  {n}^{\mu}_i {n}^{\mu}_j  - \sum_{i} B^{\mu} {n}^{\mu}_i.
\label{hamil2su3}
\end{equation}

To avoid an explicit use of the structure constants ($f_{\eta \mu \nu}$), we introduce 
the operator field $\boldsymbol{\mathfrak{n}}_j = n^{\mu}_j {\lambda}_{\mu}$.
Topological soliton solutions of the color field become well-defined in the continuum (long wavelength) limit, where the Hamiltonian can be approximated by
\begin{equation}
\mathcal{H}  \simeq  \int \mathrm{d} r^{2}\left[-\frac{{\cal I}^{\mu}_1}{2}\partial_{a} { n}^{\mu}  \partial_{a} {n}^{\mu}+\frac{{\cal I}^{\mu}_2}{2}\left(\nabla^{2} {n}^{\mu}\right)^{2}- \mathcal{B}^{\mu}  n^{\mu} \right].
\label{hamilc4}
\end{equation}
Here $\partial_a$ denotes the spatial derivatives and there is an implicit summation over the repeated $a$ index. The coupling constants can be expressed in terms of the parameters of the lattice model \eqref{hamil2}:
\begin{eqnarray}
{\cal I}^{\mu}_1 &=& \frac{3 }{2} ({I}^{\mu}_1 + 3 {I}^{\mu}_2  ),
\quad
{\cal I}^{\mu}_2 = \frac{3 }{32}(I^{\mu}_1 +9 I^{\mu}_2 ),
\nonumber \\
{\cal B}^{\mu} &=& B^{\mu} - 3(\Delta-1) (J_1+J_2) \delta_{\mu,8}.
\end{eqnarray}
Eq.~\eqref{hamilc4} corresponds to an anisotropic CP$^2$  model. 
For skyrmion solutions  the base plane $\mathbb{R}^{2}$ can be compactified to $S^2$  because the color field takes a constant value
$n_{\infty}$ at spatial infinity. These spin textures can then be characterized by the topological charge of the mapping
 $\boldsymbol{\mathfrak{n}}: \mathbb{R}^{2} \sim S^{2} \mapsto C P^{2}$:
\begin{equation}
C=-\frac{i}{32 \pi} \int \mathrm{d}x \mathrm{d}y  \varepsilon_{j k} \operatorname{Tr}\left(\boldsymbol{\mathfrak{n}}\left[\partial_{j} \boldsymbol{\mathfrak{n}}, \partial_{k} \boldsymbol{\mathfrak{n}}\right]\right).
\label{eq:skxcharge}
\end{equation}
For the lattice systems of interest, the CP$^2$ skyrmion charge can be computed after interpolating the color fields on nearest-neighbor sites $\boldsymbol{\mathfrak{n}}_j$ and  $\boldsymbol{\mathfrak{n}}_k$ along the CP$^2$ geodesic:
\begin{equation}
C=\sum_{\triangle_{j k l}} \rho_{j k l}  = 
\frac{1}{2 \pi} \sum_{\triangle_{j k l}} \left(\gamma_{j l}+\gamma_{l k}+\gamma_{k j}\right),
\label{eq:totsky}
\end{equation}
where $\triangle_{j k l}$ denotes each oriented triangular plaquette of nearest-neighbor sites $ j \to k \to l$
and 
$
\gamma_{k j}=\arg \left[\left\langle \bm{Z}_{k} \mid \bm{Z}_{j}\right\rangle\right]
$
is the Berry connection on the bond $j \to k$~\cite{supp}.
\begin{figure}[t!]
\centering
\includegraphics[width=0.48\textwidth]{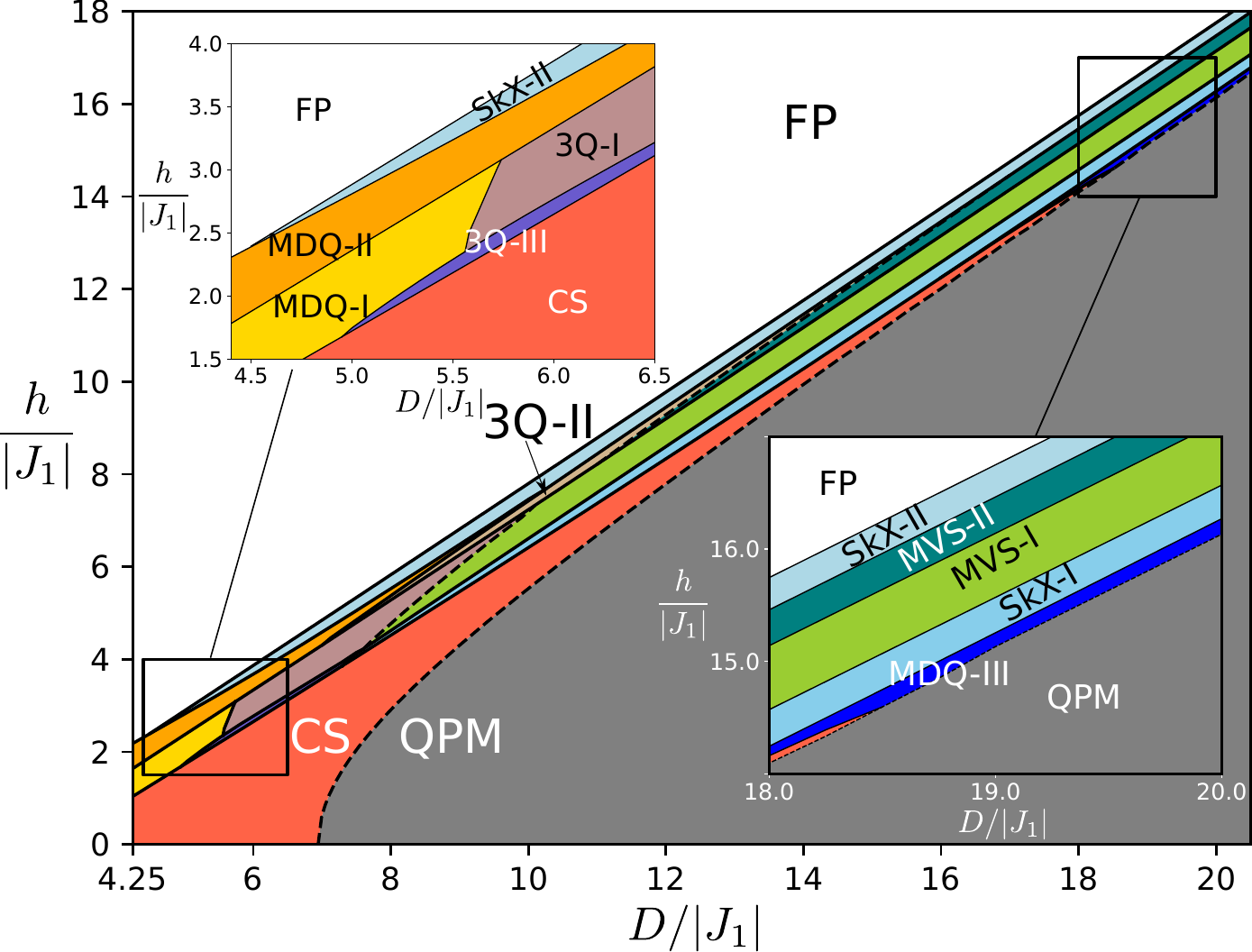}
\caption{$T=0$ phase diagram of the classical Hamiltonian ${\mathcal H}$ as a function of the single-ion anisotropy $D$ and the external field $h$, for  $J_2/|J_1| = 2/(1+\sqrt{5})$ and $\Delta=2.6$. The two insets show the phases for small-$D$ and large-$D$, respectively.  The solid (dashed) lines indicate 1st- (2nd-) order phase transitions.}
\label{fig:phasediagram}
\end{figure}

\begin{figure}[t!]
\centering
\includegraphics[width=0.48\textwidth]{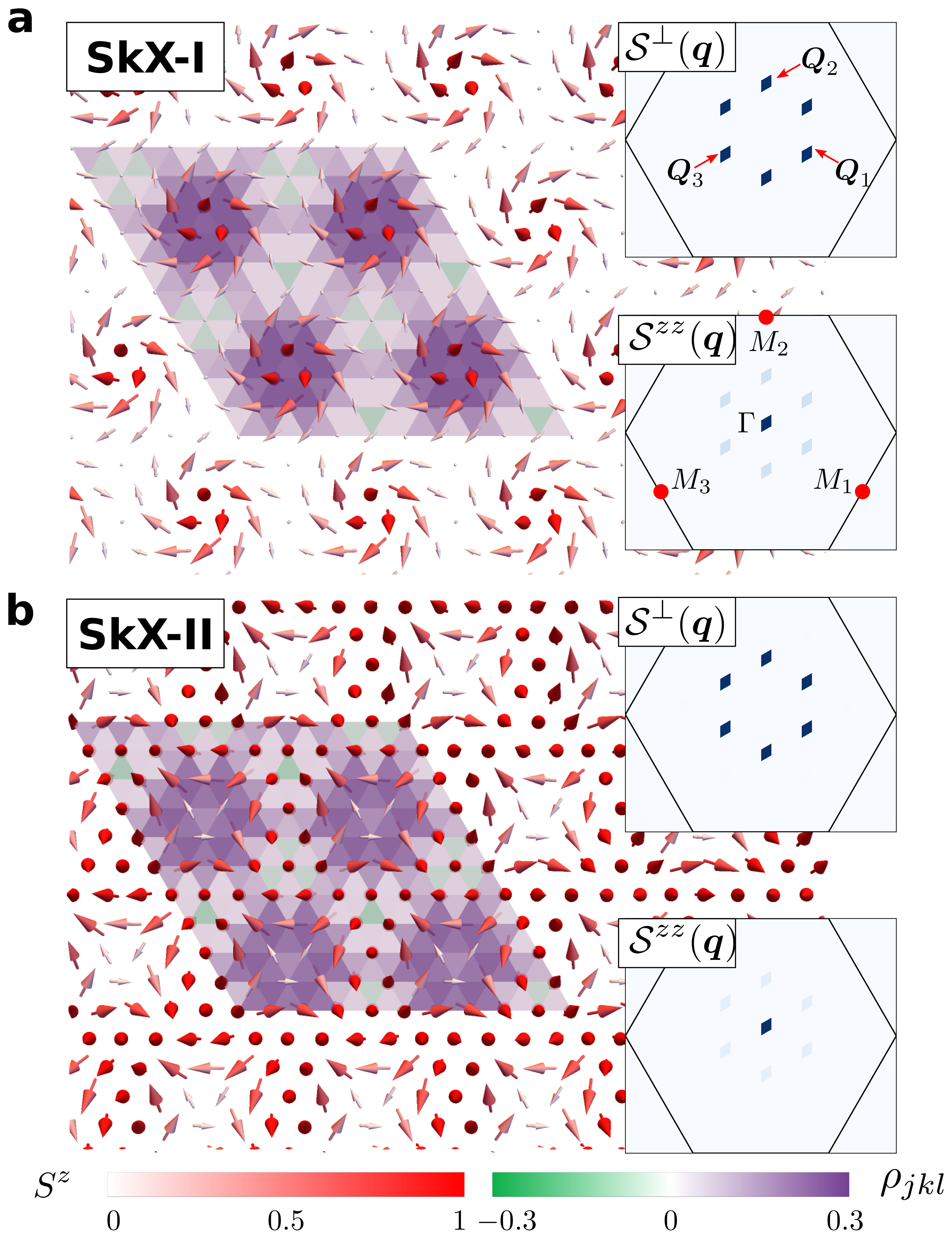}
\caption{{\bf a}, {\bf b} Real space distribution of the dipolar sector of the CP$^2$ skyrmion crystals SkX-I and SkX-II. The length of the arrow represents the magnitude of the dipole moment of the color field $|\langle \hat{\bm{S}}_j \rangle|=\sqrt{(n_j^7)^2+(n_j^5)^2+(n_j^2)^2}$. The color scale of the arrows indicates $\langle \hat{S}^z_j \rangle= -n_j^2$. The insets display the static spin structure factors $\mathcal{S}^{\perp}(\bm{q})= n^7_{\bm{q}}n^7_{\bar{\bm{q}}}+n^5_{\bm{q}}n^5_{\bar{\bm{q}}}$ and $\mathcal{S}^{zz}(\bm{q})= n^2_{\bm{q}}n^2_{\bar{\bm{q}}}$, with $\bm{n}_{\bm{q}}= \sum_{j} e^{\iu \bm{q} \cdot \bm{r_j}} \bm{n}_j/L$.  The CP$^2$ skyrmion density distribution $\rho_{j k l}$ [see Eq.~\eqref{eq:totsky}] is indicated by the color of the triangular plaquettes}.
\label{fig:SkX}
\end{figure}


\section{ Phase diagram}

The $T=0$ phase diagram is obtained by numerically minimizing the classical spin Hamiltonian $\mathcal{H}$ \eqref{hamil2su3} in the $4L^2$-dimensional phase space of a magnetic cell of $L \times L$ spins (see Methods). The shape and size of this unit cell is dictated by the symmetry-related magnetic ordering wave vectors $\bm{Q}_{\nu}$ ($\nu=1,2,3$) [see Figs.~\ref{fig:SkX}{\bf a} and {\bf b}], which are determined by minimizing the exchange interaction in momentum space:
$
J({\bm{q}}) = \sum_{jl} J_{jl} e^{\iu \bm{q} \cdot (\bm{r}_j-\bm{r}_l)}.
$ 
 The ratio between both exchange interactions, $J_2/|J_1| = 2/(1+\sqrt{5})$, is tuned to fix the magnitude of the ordering wave vectors, $|\bm{Q}_{\nu}|=|\bm{b}_1|/5$~\cite{Hayami16}, corresponding to a magnetic unit  cell of linear size $L=5$. 
A we will see later, the relevant qualitative aspects of the phase diagram do not depend on the particular choice of the model.
The three  ordering wave vectors, which are related by the $C_6$ symmetry of the TL, are parallel to the $\Gamma$-M$_\nu$ directions (denoted in Fig.~\ref{fig:SkX}).

The resulting phase diagram shown in Fig.~\ref{fig:phasediagram} includes multiple magnetically ordered phases between the FP phase and the QPM phase, where every spin is in the $| 0 \rangle$ state. For $D \gg |J_1| $, these phases include two field-induced CP$^2$ skyrmion crystals (SkX-I and SkX-II), separated by two modulated vertical spiral phases (MVS-I and MVS-II), where the polarization plane of the spiral is parallel to the $c$-axis and the magnitude of the dipole moment is continuously suppressed as the moment rotates from $\hat{\bm{z}}$ to $-\hat{\bm{z}}$ directions. The spiral phases have the same symmetry and are separated by a first order metamagnetic transition.
As shown in Fig.~\ref{fig:SkX}{\bf a}, the CP$^2$ skyrmions of the SkX-I crystal have dipole moments that evolve continuously into the purely nematic state ($| 0 \rangle$) as they move away from the core. Conversely, Fig.~\ref{fig:SkX}{\bf b}
shows that the spins in the SkX-II phase have a strong quadrupolar character (the small dipolar moment is completely suppressed in the large $D/|J_1|$ limit) at the skyrmion core, and evolve continuously into the magnetic state $| 1 \rangle$ as they move away from the core. The CP$^2$ skyrmion density distribution $\rho_{j k l}$  is also indicated with colored triangular plaquettes in Fig.~\ref{fig:SkX}{\bf a}, {\bf b} for  SkX-I and SkX-II, respectively. As shown in the inset of Fig.~\ref{fig:phasediagram}, the phase SkX-II extends down to $D/|J_1| \simeq 5$, while the phase SkX-I disappears near $D/|J_1|\simeq 8$.

New competing  orderings  appear in intermediate $D/|J_1|$ region. In particular, a significant fraction of the phase diagram is occupied by the so-called canted spiral (CS) phase, 
\begin{equation}
|\bm{Z}_{j} \rangle =\cos\theta|0\rangle_j +e^{i\bm{Q}\cdot\bm{r}_j}\sin\theta\cos\phi|1\rangle_j +e^{-i\bm{Q}\cdot\bm{r}_j}\sin\theta\sin\phi|\bar{1}\rangle_j,
\label{eq:mfs}
\end{equation}
where $\theta$ and $\phi$ are variational parameters, and $\bm{Q}$ can take any values among \{$\bm{Q}_1$, $\bm{Q}_2$, $\bm{Q}_3$\}. Upon increasing $D$, the magnitude of the dipole moment of each spin, $|\langle \hat{\bm{S}}_j \rangle|$, is continuously suppressed to zero at the boundary, 
\begin{equation}
D_c = h \sqrt{1-\frac{4 J^{2}(\boldsymbol{Q})}{h^{2}+4 J^{2}(\boldsymbol{Q})}}-2 J(\boldsymbol{Q})\left(1-\frac{2 J(\boldsymbol{Q})}{\sqrt{h^{2}+4 J^{2}(\boldsymbol{Q})}}\right),
\end{equation}
that signals the second order transition into the QPM phase. As shown in Fig.~\ref{fig:phasediagram}, several competing phases appear above the CS phase upon increasing $h$. These phases include three triple-$\bm{Q}$ spiral orderings [3$\bm{Q}$ I-III] with dominant weight in one of three $\bm{Q}$ transverse components and a staggered distribution of the CP$^2$ skyrmion density $\rho_{j k l}$ [see Eq.~\eqref{eq:totsky}] and three different modulated double-$\bm{Q}$ orderings (MDQ I - III) and two triple-$\bm{Q}$ orderings. All of these phases  are described in detail in the supplementary information. In the rest of the paper we will focus on  the SkX phases and the single-skyrmion metastable solutions that emerge in their proximity. 
\begin{figure}[t!]
\centering
\includegraphics[scale=.75]{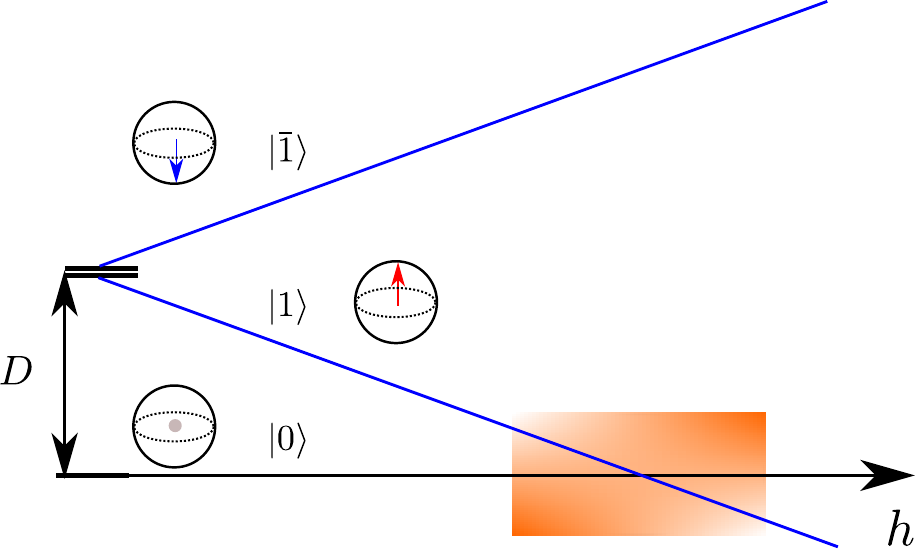}
\caption{Spectrum of the single-ion model $\hat{\mathcal{H}}_\text{SI}= D(\hat{S}^z)^2 - h \hat{S}^z$. 
The shaded region denotes the effective regime with a quasi-degenerate doublet: $\{|0\rangle ,|1\rangle \}$ }.
\label{fig:SionSpec}
\end{figure}

\section{Large-$D$ limit}

The origin of the CP$^2$ skyrmion crystals can be understood by analyzing the small $|J_{ij}|/D$ regime, where
$\hat{\mathcal{H}}$ can be reduced via first order degenerate perturbation theory in $J_{ij}/D$ to an effective pseudo-spin-$1/2$ low-energy Hamiltonian,
\begin{equation}
\hat{\mathcal{H}}_\text{eff}= \sum_{\langle i,j \rangle} {\tilde J}_{ij} ( \hat{s}^x_i \hat{s}^x_j + \hat{s}^y_i \hat{s}^y_j + \tilde{\Delta} \hat{s}^z_i \hat{s}^z_j) - {\tilde h} \sum_{i}  \hat{s}^z_i.
\label{hamilsu2}
\end{equation}
The pseudo-spin-$1/2$ operators are the projection of the original spin operators into the low-energy subspace
 ${\cal S}_0$ generated by the quasi-degenerate  doublet $\{ |0 \rangle_j,  | 1\rangle_j \}$ (see Fig.~\ref{fig:SionSpec}):
\begin{equation}
\hat{s}^z_j = {\cal P}_0 \hat{S}^z_j {\cal P}_0 - \frac{1}{2},
\quad
\hat{s}^{\pm}_j = \frac{ {\cal P}_0 \hat{S}^{\pm}_j {\cal P}_0}{\sqrt{2}},
\label{pseudo}
\end{equation}
where ${\cal P}_0$ is the projection operator of the low-energy subspace.
Importantly, the first state of the doublet has a net quadrupolar moment but no net dipole moment,
$ \langle 0| \hat{\bm{S}}_j |0 \rangle_j=0 $, while the second state maximizes the  dipole moment along the $\hat{\bm{z}}$-direction
$ \langle 1 | \hat{\bm{S}}_j |1 \rangle_j= \hat{\bm{z}} $. This means that three pseudo-spin operators 
 generate an SU(2) subgroup of SU(3) different from the SU(2) subgroup of spin rotations.

$\hat{\mathcal{H}}_\text{eff}$ represents an effective triangular easy-axis XXZ model with effective exchange, anisotropy and field parameters  ${\tilde J}_{ij} = 2 J_{ij}$, $\tilde{\Delta} = \frac{\Delta}{2}$ and ${\tilde h} = h - D - 3  \Delta (J_1 + J_2 )$, respectively. This model is known to exhibit a field-induced CP$^1$ SkX phase~\cite{Leonov2015,Hayami16} that survives in the long-wavelength limit~\cite{Lin2016_skyrmion}:
\begin{equation}
\mathcal{H}_\text{eff}  \simeq \int \mathrm{d} r^{2}\left[-\frac{{\cal J}^{\eta}_1}{2}\partial_{a}\tilde{ n}^{\eta} \cdot \partial_{a} \tilde{n}^{\eta}+\frac{{\cal J}^{\eta}_2}{2}\left(\nabla^{2} \tilde{n}^{\eta}\right)^{2}- \tilde{\cal B}  \tilde{n}_z + \tilde{\cal D} \tilde{n}_{z}^{2}  \right],
\label{hamil3}
\end{equation}
where   $\eta=x, y, z$ denotes the three  components of the unit vector field $\tilde{\boldsymbol n} $  ($\vert \tilde{\boldsymbol n} \vert=1$), and
\begin{eqnarray}
{\cal J}^{\eta}_1 &=& \frac{3 s^2}{2} (\tilde{J}_1 + 3 \tilde{J}_2) [1+ (\tilde{\Delta}-1) \delta_{\eta z}],
\nonumber \\
{\cal J}^{\eta}_2 &=& \frac{3 s^2}{32}(\tilde{J}_1+9 \tilde{J}_2) [1+ (\tilde{\Delta}-1) \delta_{\eta z}]
\nonumber \\
\tilde{\cal B} &=& s {\tilde h}, 
\quad \tilde{\cal D} = 3 s^2 (\tilde{\Delta}-1) (\tilde{J}_1 +\tilde{J}_2),
\end{eqnarray}
where $s=1/2$.
Although the target manifold of this theory is CP$^1$ (orbit of SU(2) coherent states that belong ${\cal S}_0$), we must keep in mind that $\hat{\mathcal{H}}_\text{eff}$ describes
the large $D/|J_1|$ limit where the CP$^2$ skyrmions of the original spin-$1$ model become asymptotically close to CP$^1$ \emph{pseudo-spin} skyrmions. In other words, the SkXs include a finite $|\bar{1} \rangle$ component for finite $D/|J_1|$, which increases upon decreasing $D$. This component, which only appears in the low-energy model when second order corrections in $J_{ij}$ are included, is responsible for the metamagnetic transition between the MVS-I and MVS-II phases (the transition disappears in the $D \to \infty$ limit). 

\begin{figure}[t!]
\centering
\includegraphics[width=.48\textwidth]{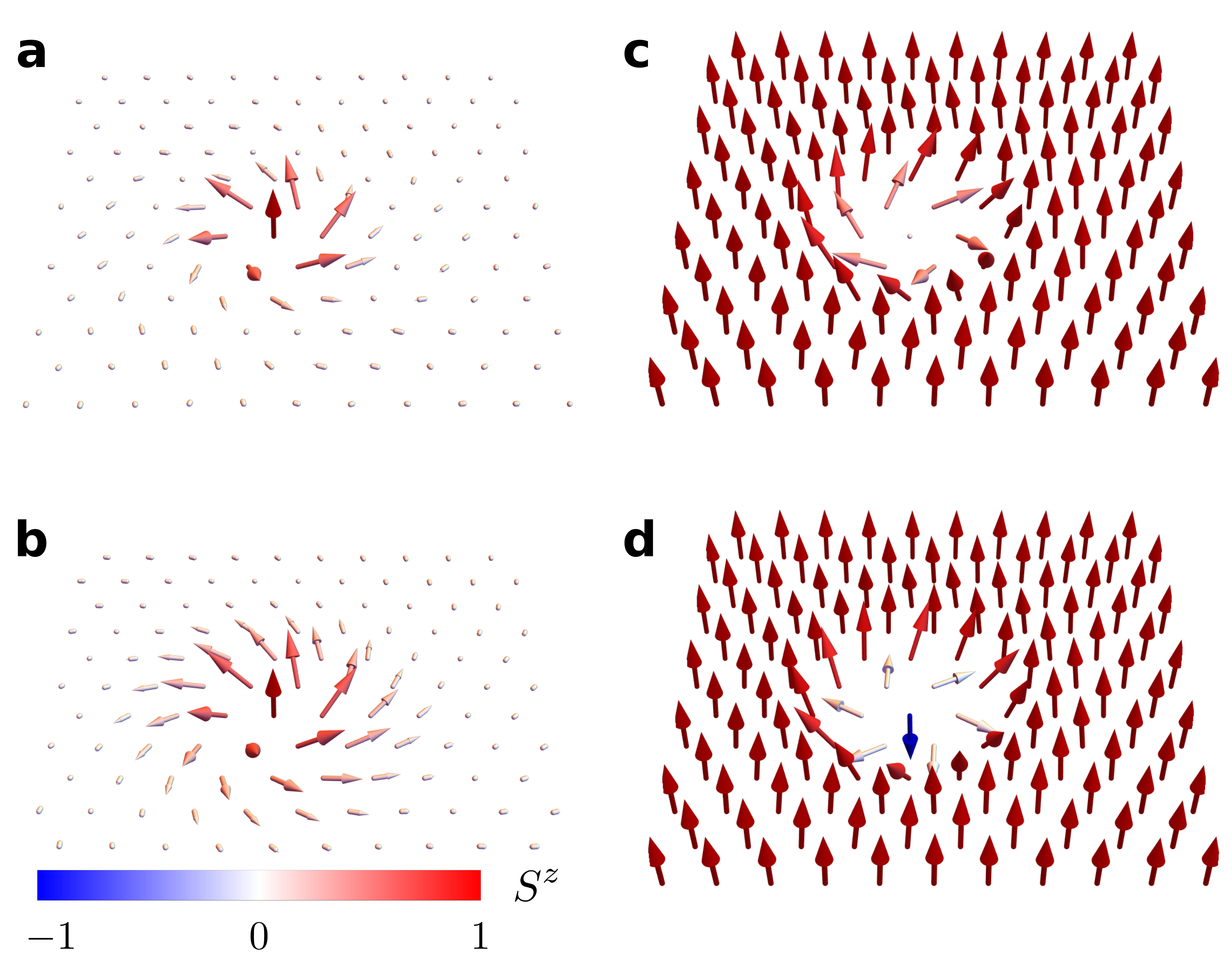}
\caption{Dipolar sector of CP$^2$ skyrmions. The color scale indicates the value of $n_j^2$ ($\langle \hat{S}_j^z \rangle$). {\bf a}, {\bf b} Skyrmion excitation on top of a QPM background. {\bf c}, {\bf d} Skyrmion excitation on top of a fully polarized background. $J_2/|J_1|=2/(1+\sqrt{5})$ and $\Delta=2.6$ in (a), (c), and (d). $J_2/|J_1|=2/(3+\sqrt{5})$ and $\Delta=2.2$ in {\bf b}.  In these panels, {\bf a}. $D=17.1|J_1|$, $H=13|J_1|$. {\bf b}. $D=18.3|J_1|$, $H=14|J_1|$. {\bf c}. $D=7|J_1|$, $H=5|J_1|$. {\bf d}. $D=4|J_1|$, $H=2|J_1|$. }
\label{fig:Skyrmions}
\end{figure}

Since $\hat{\mathcal{H}}_\text{eff}(h)$ and $\hat{\mathcal{H}}_\text{eff}(-h)$ are related by a pseudo-time-reversal (PTR) transformation
($\hat{s}_j \to - \hat{s}_j$ on the lattice and $\tilde{n} \to - \tilde{n}$ in the continuum) their corresponding ground states are related by the same transformation. 
In particular, the ground state   $(\tilde{\boldsymbol n} = \hat{\bm{z}})$ that is obtained above the saturation field ($\tilde{\cal B} > \tilde{\cal B}_{\rm sat}$) corresponds to the FP state ($\langle \hat{\bm{S}}_j \rangle = \hat{\bm{z}} $) in the original spin-$1$ variables, while the ground state $(\tilde{\boldsymbol n} = -\hat{\bm{z}})$ below the negative saturation field
($ \tilde{\cal B} < -\tilde{\cal B}_{\rm sat}$) corresponds to the QPM phase ($\vert \bm{Z}_j \rangle = \vert 0 \rangle_j$).
Correspondingly, the SkX  induced by a positive $h$ has pseudo-spins  polarized along the quadrupolar  direction
($| 0 \rangle$) near the core of the skyrmions and  parallel to the dipolar one ($| 1 \rangle$) at the midpoints between two cores. This explains the origin of the SkX-II crystals depicted in Fig.~\ref{fig:SkX}~{\bf{b}}.
The negative ${\cal B}$ counterpart of this phase, which is obtained by applying the PTR transformation, corresponds to
the  SkX-I crystal shown in Fig.~\ref{fig:SkX}{\bf a}. In this case the skyrmion cores are magnetic, while  the midpoints are practically quadrupolar (they become purely quadrupolar in the large $D/|J_1|$ limit).
This simple reasoning explains the origin of the novel SkX phases included in the $T=0$ phase diagram of $\mathcal{H}$ shown in Fig.~\ref{fig:phasediagram}. The intermediate phase between the SkX-I and SkX-II ground state of
${\mathcal{H}}$  induced by positive and negative values of 
$h$ is a single-$\bm{Q}$ spiral  with a polarization plane parallel to the $c$-axis known as vertical spiral (VS). This explains the origin of 
the MVS-I and MVS-II phases in between the two SkX phases (the first order transition between both phases disappears in the large-$D$ limit~\cite{Leonov2015}).


{\it Single-skyrmion solutions.} Besides the SkX phases shown in Fig.~\ref{fig:Skyrmions}, the effective field theory \eqref{hamil3} is known to support metastable CP$^1$ single-skyrmion solutions beyond the saturation fields $ \vert \tilde{\mathcal{B}} \vert > \tilde{\mathcal{B}}_{\rm sat}$. The pseudo-spin variable is anti-parallel to the external field at the core and it gradually rotates towards the  direction parallel to the field upon moving away from the center. Interestingly, this pseudo-spin texture translates into metastable single-skyrmion solutions of the QPM phase that have a magnetic core and a nematic periphery, as it is shown in Figs.~\ref{fig:Skyrmions}{\bf a} and {\bf b} for different sets of Hamiltonian parameters.  The  CP$^2$ skyrmions  are metastable solutions in the QPM phase for $D \gtrsim 14$, implying that these exotic magnetic-nematic textures should emerge in real magnets under quite general conditions. 

Similarly, the metastable pseudo-spin single-skyrmion solutions of the FP phase ($\tilde{\mathcal{B}} > \tilde{\mathcal{B}}_{\rm sat}$) lead to a spin texture with a  nematic (non-magnetic) core and a magnetic (FP) periphery, like the one shown in Fig.~\ref{fig:Skyrmions}{\bf c}. Interestingly, this exotic CP$^2$ skyrmion solution remains metastable down to $D \simeq 4 \vert J_1 \vert$ and it coexists with regular (CP$^1$) metastable skyrmion solutions, like the one shown  Fig.~\ref{fig:Skyrmions}{\bf d}, that emerge below $D \simeq 4.25 |J_1| $.

\section{ Discussion}

In summary, we have demonstrated that  CP$^{2}$ skyrmion textures  emerge in realistic models of hexagonal magnets out of the combination of geometric frustration with competing exchange and single-ion anisotropies. It is important to note that the skyrmion crystals and metastable solutions reported in this work survive in the long wavelength limit~\cite{Lin2016_skyrmion}, implying that the above described CP$^2$ skyrmion phases should also exist in extensions of the model to honeycomb and Kagome lattice geometries. 
The generic spin-$1$ model considered in this work describes a series of 
triangular antiferromagnets in the form of ABX$_3$, BX$_2$, and ABO$_2$~\cite{Collins1997_review,McGuire17,Botana19}, where A is a an alkali metal, B is a transition metal, and X is a halogen atom.
Several Ni-based compounds, including NiGa$_2$S$_4$~\cite{Nakatsuji2005}, Ba$_3$NiSb$_2$O$_9$~\cite{Cheng2011}, Na$_2$BaNi(PO$_4$)$_2$~\cite{Li2021}, are also found to be realizations of spin-$1$ models on TLs.
Some of these compounds have been already identified as candidates to host crystals of
CP$^1$ magnetic skyrmions that are stabilized by the combination of frustration and exchange anisotropy~\cite{Amoroso20}. Others, such as FeI$_2$~\cite{Bai21,Legros21}, are described by the the Hamiltonian \eqref{hamil} but the sign of the single-ion and exchange anisotropies is opposite to the case of interest in this work.
Related compounds, such as CsFeCl$_3$, are  known to be quantum paramagnets described by the Hamiltonian \eqref{hamil}~\cite{Kurita16} with a dominant easy-plane single-ion anisotropy $D/J_1 \simeq 10$. 
An alternative route to find realizations of our spin-$1$ Hamiltonian is to consider hexagonal materials  comprising  $4f$ magnetic ions with a singlet single-ion ground state and an excited Ising-like doublet (the effective easy-plane single-ion anisotropy $D$ is equal to the singlet-doublet gap).
Ultracold atoms are also powerful platforms to realize spin-$1$ models with {\it tunable} single-ion anisotropy~\cite{Chung2021}.

Our results demonstrate that magnetic field induced CP$^{2}$ skyrmion crystals should emerge in the presence of a dominant single-ion easy-plane anisotropy $D$ that is strong enough to stabilize a QPM at $T=0$.
In terms of SU(3) spins, the single-ion anisotropy acts as an external SU(3) field that couples linearly to one component (${\hat T}_j^{8}$) of the quadrupolar moment. Correspondingly, the 
QPM can be regarded as a uniform quadrupolar state induced by a strong enough ${\hat T}_j^{8}$ component of the SU(3) field. The field-induced quantum phase transition between this uniform quadrupolar state and the skyrmion crystals is
presaged by the emergence of metastable CP$^2$ single-skyrmion solutions consisting  of a magnetic skyrmion core that decays continuously into a quadrupolar periphery. These novel skyrmions  can be induced by increasing the strength of the magnetic field in the neighborhood of a given magnetic ion of a frustrated quantum paramagnet with competing exchange and single-ion anisotropies. 

\section{Acknowledgments}

We acknowledge useful discussions with Xiaojian Bai, Antia Botana, Ying Wai Li, Shizeng Lin, Cole Miles, Martin Mourigal, Sakib Matin, Matthew Wilson, and Shang-Shun Zhang.
D. D., K. B. and C.D.B.~acknowledge support from U.S. Department of Energy, Office of Science, Office of Basic Energy Sciences, under Award No.~DE-SC0022311.  The work by H.Z. was supported by the Graduate Advancement, Training and Education (GATE) fellowship.
Z.W. was supported by the U.S. Department of Energy through the University of Minnesota Center for Quantum Materials, under Award No.~DE-SC-0016371.

\section{Methods}

The numerical minimization for the phase diagram Fig.~\ref{fig:phasediagram} is done in a cell of $10 \times 10$ spins containing four magnetic unit cells ($L=5$). Two crucial steps are useful to improve the efficiency of the local gradient-based minimization algorithms~\cite{nlopt}. In the first step, we set multiple random initial conditions $|\bm{Z}\rangle$ ($\sim$ 50 for our case), where $|\bm{Z}_j\rangle$ on \emph{every} site $j$ is uniformly sampled on the CP$^2 \simeq S^5/S^1$ manifold. After running the minimization algorithm, we keep the solution with the lowest energy for a given set of Hamiltonian parameters. In the next step, half of the initial conditions are randomly generated, while the other half correspond to the lowest-energy solutions obtained in the first step within a predefined neighborhood of the  Hamiltonian parameters. This procedure is iterated until the phase diagram converges.

\bibliographystyle{naturemag}

\end{document}


\title{Supplemental Information for ``\texorpdfstring{$\text{CP}^2$}{CP2} Skyrmions and Skyrmion Crystals in Realistic Quantum Magnets''}
\maketitle

\author{Hao~Zhang}
\affiliation{Department of Physics and Astronomy, The University of Tennessee,
Knoxville, Tennessee 37996, USA}
\affiliation{Materials Science and Technology Division, Oak Ridge National Laboratory, Oak Ridge, Tennessee 37831, USA}
\author{Zhentao~Wang}
\affiliation{Department of Physics and Astronomy, The University of Tennessee,
Knoxville, Tennessee 37996, USA}
\affiliation{School of Physics and Astronomy, University of Minnesota, Minneapolis, Minnesota 55455, USA}
\author{David~Dahlbom}
\affiliation{Department of Physics and Astronomy, The University of Tennessee,
Knoxville, Tennessee 37996, USA}
\author{Kipton~Barros}
\affiliation{Theoretical Division and CNLS, Los Alamos National Laboratory, Los Alamos, New Mexico 87545, USA}
\author{Cristian~D.~Batista}
\affiliation{Department of Physics and Astronomy, The University of Tennessee,
Knoxville, Tennessee 37996, USA}
\affiliation{Quantum Condensed Matter Division and Shull-Wollan Center, Oak Ridge
National Laboratory, Oak Ridge, Tennessee 37831, USA}

\setcounter{figure}{0}
\global\long\def\thefigure{S\arabic{figure}}%
\setcounter{table}{0}
\global\long\def\thetable{S\arabic{table}}%
\setcounter{equation}{0}
\global\long\def\theequation{S\arabic{equation}}%

\section{Characterization of phases}
In the main text we present detailed discussions for the two skyrmion crystal phases (SkX-I and SkX-II) that emerge from the classical SU(3) limit of the model Hamiltonian [see Eq.~(12) of the main text]. Here we characterize the remaining phases reported in Fig.~1 of the main text.

\subsection{Quantum paramagnet phase and fully-polarized state}
The classical limit of the quantum paramagnet (QPM) is described by the coherent state $|\bm{Z}_j\rangle = |0\rangle_j$, for which all three dipolar components of the color field vanish: $\langle \bm{Z}_j| \hat{S}^{x,y,z}_j | \bm{Z}_j \rangle=0$. However, the remaining five nematic components of the color field may take non-zero values, e.g. $\langle \bm{Z}_j| \hat{T}^8_j |\bm{Z}_j \rangle = 1/\sqrt{3}$. In contrast, the fully-polarized state is described by the coherent state $|\bm{Z}_j\rangle = |1\rangle_j$ that maximizes the dipolar moment along the $z$ axis: $\langle \bm{Z}_j | \hat{S}^z_j| \bm{Z}_j\rangle=1$.

\subsection{Single-\texorpdfstring{$\bm{Q}$}{Q} orderings}
\begin{figure}
    \centering
    \includegraphics[width=.95\textwidth]{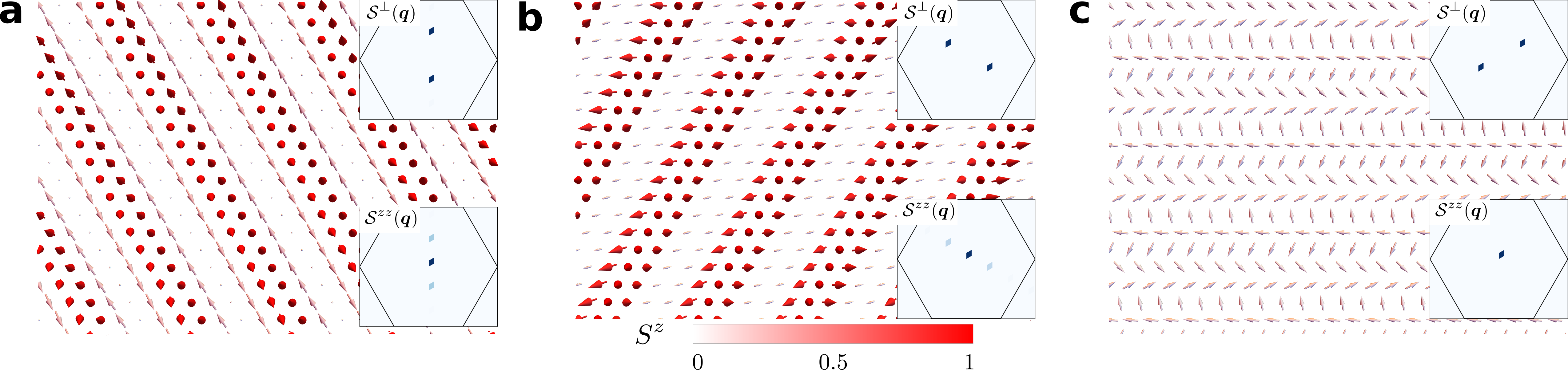}
    \caption{Real space distribution of the dipolar sector of the color fields for the three single-$\bm{Q}$ orderings. {\bf a} MVS-I. {\bf b} MVS-II. {\bf c} CS. The length of the arrow represents the magnitude of the dipole moment of the color field $|\langle \hat{\bm{S}}_j \rangle|=\sqrt{(n_j^7)^2+(n_j^5)^2+(n_j^2)^2}$. The color scale of the arrows indicates  $\langle \hat{S}^z_j \rangle = -n_j^2$. The insets display the static spin structure factors $\mathcal{S}^{\perp}(\bm{q})=\langle n^7_{\bm{q}}n^7_{\bar{\bm{q}}}+n^5_{\bm{q}}n^5_{\bar{\bm{q}}}\rangle$ and $\mathcal{S}^{zz}(\bm{q})=\langle n^2_{\bm{q}}n^2_{\bar{\bm{q}}}\rangle$, with $\bm{n}_{\bm{q}}= \sum_{j} e^{\iu \bm{q} \cdot \bm{r_j}} \bm{n}_j/\sqrt{N}$.}
    \label{fig:mvs1QS}
\end{figure}

A large region of the phase diagram is occupied by the canted spiral (CS) phase described by the coherent state
\begin{equation}
|\bm{Z}_{j} \rangle =\cos\theta|0\rangle+e^{i\bm{Q}\cdot\bm{r}_j}\sin\theta\cos\phi|1\rangle+e^{-i\bm{Q}\cdot\bm{r}_j}\sin\theta\sin\phi|\bar{1}\rangle,
\end{equation}
where $\theta$ and $\phi$ are variational parameters and $\bm{Q}$ is the ordering wave vector. Note that the CS phase has an uniform value of $\langle \hat{S}_j^z \rangle$ for all sites [see Fig.~\ref{fig:mvs1QS}{\bf c}]. For a fixed external magnetic field $h$, the dipole moment of each site, $\langle \bm{S}_j \rangle$, is continuously suppressed to zero at
\begin{equation}
D_c = h \sqrt{1-\frac{4 J^{2}(\boldsymbol{Q})}{h^{2}+4 J^{2}(\boldsymbol{Q})}}-2 J(\boldsymbol{Q})\left(1-\frac{2 J(\boldsymbol{Q})}{\sqrt{h^{2}+4 J^{2}(\boldsymbol{Q})}}\right),
\end{equation}
that signals the second order transition into the QPM phase.

In between the two Skyrmion crystal phases, there are two modulated vertical spiral phases MVS-I and MVS-II with a polarization plane parallel to the $\hat{\bm{z}}$-axis. Unlike the case of spirals described by SU(2) coherent states, the magnitude of the dipole moment is continuously modulated as the dipole moment rotates along the $\hat{\bm{z}}$ axis [see Fig.~\ref{fig:mvs1QS}{\bf a}~{\bf b}]. In the large-$D/|J_1|$ limit, the two spirals become vertical spiral states in the pseudo-spin variables. 

\subsection{Modulated double-\texorpdfstring{$\bm{Q}$}{Q} orderings}
\begin{figure}[t!]
    \centering
    \includegraphics[width=.95\textwidth]{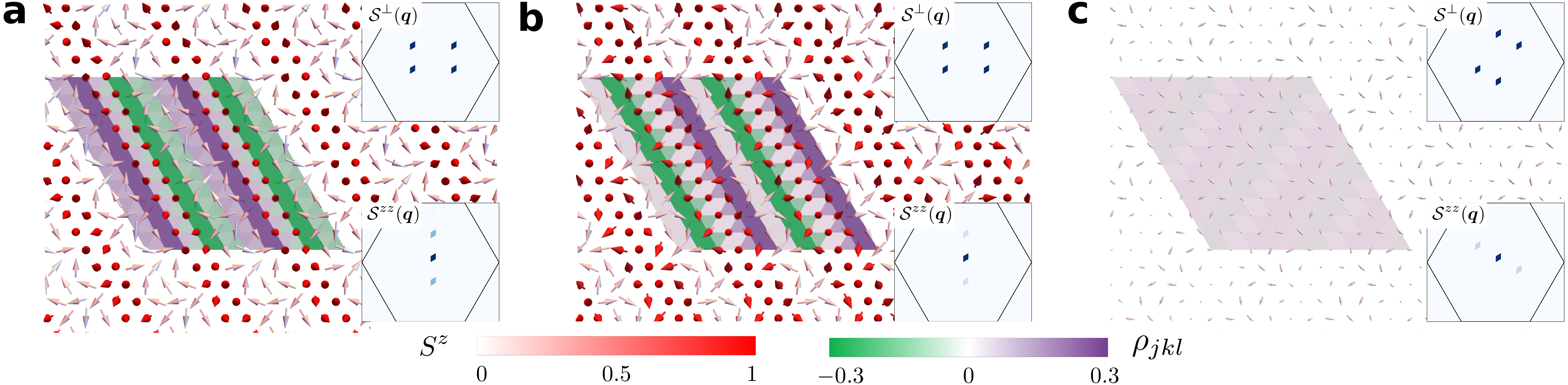}
    \caption{Real space distribution of the dipolar sector of the color fields for the three modulated double-$\bm{Q}$ orderings: {\bf a} MDQ-I, {\bf b} MDQ-II, and {\bf c} MDQ-III. The length of the arrow represents the magnitude of the dipole moment of the color field $|\langle \hat{\bm{S}}_j \rangle|=\sqrt{(n_j^7)^2+(n_j^5)^2+(n_j^2)^2}$. The color scale  of the arrows indicates $\langle \hat{S}^z_j \rangle= -n_j^2$. The insets display the static spin structure factors $\mathcal{S}^{\perp}(\bm{q})=\langle n^7_{\bm{q}}n^7_{\bar{\bm{q}}}+n^5_{\bm{q}}n^5_{\bar{\bm{q}}}\rangle$ and $\mathcal{S}^{zz}(\bm{q})=\langle n^2_{\bm{q}}n^2_{\bar{\bm{q}}}\rangle$, with $\bm{n}_{\bm{q}}= \sum_{j} e^{\iu \bm{q} \cdot \bm{r_j}} \bm{n}_j/\sqrt{N}$.  The CP$^2$ skyrmion density distribution $\rho_{j k l}$ [see Eq.~(16) of the main text] is indicated by the color of the triangular plaquettes in all three panels.}
    \label{fig:mdq}
\end{figure}

There are three different modulated double-$\bm{Q}$ (MDQ I-III) orderings in the phase diagram [see Fig.~\ref{fig:mdq}]. Similar to the relation between MVS-I and MVS-II, the MDQ-I and MDQ-II phases appearing in the small $D/|J_1|$ region have the same symmetry and are separated by a first order metamagnetic transition. The MDQ-III phase, which occupies a small region above the QPM phase in the large $D/|J_1|$ region, is the pseudospin counterpart  of the 2-$q'$ phase reported in Ref.~\cite{Leonov2015}. Since we have chosen a strong enough easy-axis exchange anisotropy $\Delta$ [see Eq.~(1) of the main text], which translates into an effective single-ion easy-axis anisotropy $\tilde{\cal D}$ ($K$ term of  Ref.~\cite{Leonov2015}) upon taking the long wavelength limit of the effective pseudospin model [see Eq.~(22) of the main text],  the MDQ-III phase eventually disappears for $D/|J_1| \gtrsim 25 $. The existence of the MDQ-III phase for moderately large values of $D/|J_1|$ results from higher order terms in $J_{ij}/D$ not included in the effective pseudospin model (19) of the main text. As shown in Fig.~\ref{fig:mdq}, the CP$^2$ skyrmion charge distribution of these MDQ orderings displays a stripe structure.

\subsection{3\texorpdfstring{$\bm{Q}$}{Q} spiral orderings}
\begin{figure}[t!]
    \centering
    \includegraphics[width=.95\textwidth]{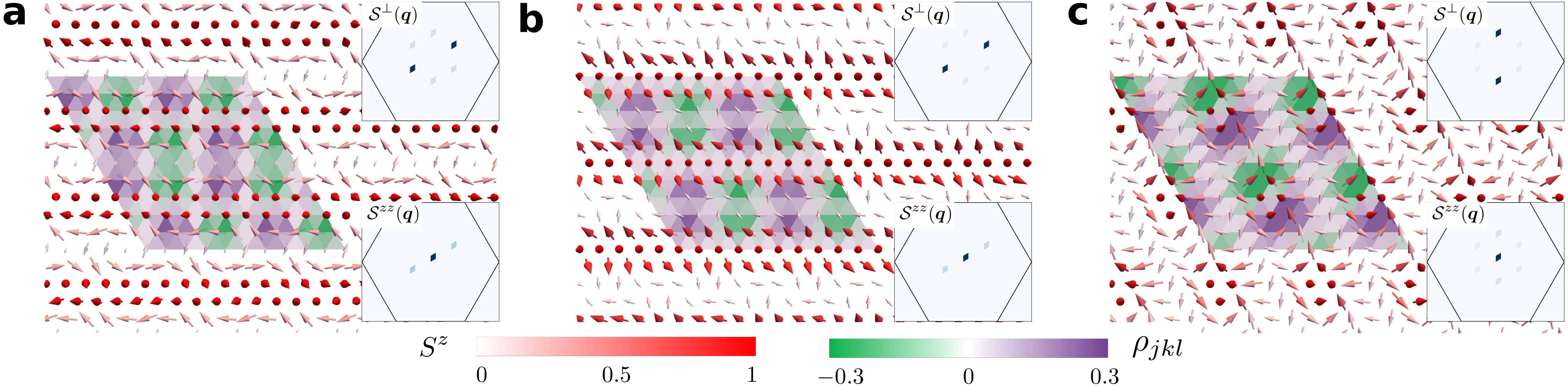}
    \caption{Real space distribution of the dipolar sector of the color fields for the three 3$\bm{Q}$ spiral orderings: {\bf a} 3$\bm{Q}$S-I, {\bf b} 3$\bm{Q}$S-II, and {\bf c} 3$\bm{Q}$S-III. The length of the arrow represents the magnitude of the dipole moment of the color field $|\langle \hat{\bm{S}}_j \rangle|=\sqrt{(n_j^7)^2+(n_j^5)^2+(n_j^2)^2}$. The color scale of the arrows indicates  $\langle \hat{S}^z_j \rangle= -n_j^2$. The insets display the static spin structure factors $\mathcal{S}^{\perp}(\bm{q})=\langle n^7_{\bm{q}}n^7_{\bar{\bm{q}}}+n^5_{\bm{q}}n^5_{\bar{\bm{q}}}\rangle$ and $\mathcal{S}^{zz}(\bm{q})=\langle n^2_{\bm{q}}n^2_{\bar{\bm{q}}}\rangle$, with $\bm{n}_{\bm{q}}= \sum_{j} e^{\iu \bm{q} \cdot \bm{r_j}} \bm{n}_j/\sqrt{N}$.  The CP$^2$ skyrmion density distribution $\rho_{j k l}$ [see Eq.~(16) of the main text] is indicated by the color of the triangular plaquettes in all three panels.}
    \label{fig:3qs}
\end{figure}

There are three triple-$\bm{Q}$ spiral orderings [3$\bm{Q}$S I-III] whose transverse spin structure factor exhibits dominant weight in one of the three ordering wave vectors $\bm{Q}_{\nu}$ ($\nu=1,2,3$) [see Fig.~\ref{fig:3qs}~{\bf a}-{\bf c}]. The CP$^2$ Skyrmion density of these 3$\bm{Q}$S phases displays a staggered distribution. Upon increasing $D$ for a fixed value of $h$ the subdominant weights are continuously suppressed for the 3$\bm{Q}$S-I and 3$\bm{Q}$S-II states, leading to a second order phase transition into MVS-I, and MVS-II, respectively. The additional characteristic of the  3$\bm{Q}$S-III state is that the longitudinal spin structure factor  has a equal weights on the three ordering wave vectors $\bm{Q}_{\nu}$.

\section{Berry phase and solid angle in \texorpdfstring{$\textbf{CP}^2$}{CP2}}
\subsection{Continuum limit}
Topological soliton solutions of the classical color field become well-defined in the continuum limit that holds for $\lambda \gg a$,
where $\lambda$ is the characteristic wave-length of the spin configuration and $a$ is the lattice parameter. In this limit, we use a Taylor
expansion to express the value of the field $n_{j}^{m}$ on the site $j$ (with coordinates $\bm{r}_{j}$ ) in terms of the field $n_{i}^{m}$
on the neighboring site located at $\bm{r}_{i}$:
\begin{equation}
n_{j}^{m}=n_{i}^{m}+r_{i j} \partial_{k} n_{i}^{m}+O\left(r_{i j}^{2}\right)
\label{eq:contl}
\end{equation}
where $\bm{r}_{j}=\bm{r}_{i}+r_{i j} \boldsymbol{e}_{k}, r_{i j}=\left|\bm{r}_{i}-\bm{r}_{j}\right|$ 
($r_{i j}$ coincides with the lattice parameter a when i and j are nearest-neighbor sites) and $\boldsymbol{e}_{k}$
is a unit vector that points in the $k$-direction of the plane. Since the lattice spin Hamiltonians that we will consider include only
interactions between spins on neighboring sites, we can use Eq.~\eqref{eq:contl} to define a continuum limit of the spin Hamiltonian by
replacing the sum over lattice sites with the integral $a^{-2} \int \mathrm{d}^{2} x$.

Since the color field takes a constant value
$n_{\infty}$ at spatial infinity for skyrmion configurations, the base plane $\mathbb{R}^{2}$ can be compactified to $S^2$. Thus, these spin textures are characterized by the topological degree (or skyrmion charge) of the mapping
 $\boldsymbol{\mathfrak{n}}:  S^{2} \mapsto C P^{2}$:
\begin{equation}
C =-\frac{i}{32 \pi} \int \mathrm{d}x \mathrm{d}y  \varepsilon_{j k} \operatorname{Tr}\left(\boldsymbol{\mathfrak{n}}\left[\partial_{j} \boldsymbol{\mathfrak{n}}, \partial_{k} \boldsymbol{\mathfrak{n}}\right]\right)
\label{eq:skxcharge}
\end{equation}
It is important to note that there is a one-to-one correspondence between the color field $\bm{\mathfrak{n}}_j$ and $\bm{Z}_j$ that defines a coherent
state. In other words, we can also express the Hamiltonian and the skyrmion density in terms of the $\bm{Z}_j$ field that in
the continuum limit becomes a $\bm{Z}({\bm r})$ field. We can think of both alternative descriptions as the classical limit of the Schr\"{o}dinger
and the Heisenberg representations. In the former case, the dynamical variables are wave functions that become coherent states
with coordinates $\bm{Z}_j$ . In the latter case, the dynamical variables are operators ${\hat T}^{\mu}_j$ (observables), which in the classical limit are
replaced by their expectation value $\left\langle T_{j}^{\mu}\right\rangle$ that coincides with the color field $n_{j}^{\mu}$.

Depending on the application, it may be more convenient to work with one representation of the classical SU(3) spin field
or with the other one. For this reason, it is useful to derive an expression of the skyrmion density in terms of the $\bm{Z}_j$-field. As
expected, the skyrmion density defined in Eq.~\eqref{eq:skxcharge} turns out to be proportional to the Berry curvature of the $\bm{Z}({\bm r})$  field. To
demonstrate this statement we just need to introduce the SU(3) Berry connection,
\begin{equation}
\mathcal{A}(\bm{r})=i\left\langle \bm{Z}\left|\nabla_{\bm{r}}\right| \bm{Z}\right\rangle
\end{equation}
and the corresponding SU(3) Berry curvature:
\begin{equation}
\mathcal{B}(\bm{r})=\nabla \times \mathcal{A}(\bm{r}).
\end{equation}
According to Stokes theorem, the integral of the Berry connection over a closed loop ${\cal C}$ is equal to the integral of the Berry
curvature over the oriented surface enclosed by ${\cal C}$:
\begin{equation}
\oint_{C} \mathcal{A}_{j} d x^{j}=\int_{S_{C}}\left[\partial_{x} \mathcal{A}_{y}-\partial_{y} \mathcal{A}_{x}\right] \mathrm{d}x \mathrm{d}y.
\label{eq:stokes}
\end{equation}
Our next goal is to demonstrate that the skyrmion density is proportional to the Berry curvature:
\begin{equation}
\partial_{x} \mathcal{A}_{y}-\partial_{y} \mathcal{A}_{x}=-\frac{i}{16} 2 \operatorname{Tr}\left(\boldsymbol{\mathfrak{n}}\left[\partial_{x} \boldsymbol{\mathfrak{n}}, \partial_{y} \boldsymbol{\mathfrak{n}}\right]\right)=-\frac{i}{16} \varepsilon_{j k} \operatorname{Tr}\left(\boldsymbol{\mathfrak{n}}\left[\partial_{j} \boldsymbol{\mathfrak{n}}, \partial_{k} \boldsymbol{\mathfrak{n}}\right]\right).
\label{eq:prop}
\end{equation}
To demonstrate this equivalence, we first need to demonstrate that the infinitesimal ``geodesic'' SU(3) spin rotation that transforms
the coherent state $|\bm{Z}(\bm{r})\rangle$ into $|\bm{Z}(\bm{r}+\delta \bm{r})\rangle$ up to a phase factor $e^{i \delta \varphi}$ is:
\begin{equation}
e^{i \delta \varphi}|\bm{Z}(\bm{r}+\delta \bm{r})\rangle=\hat{U}_{\bm{r}+\delta \bm{r}, \bm{r}}|\bm{Z}(\bm{r})\rangle=\left(\mathbb{1}-\frac{i}{4} f_{\mu v \eta} n^{\mu} \partial_{\bm{r}} n^{\nu} \cdot \delta \bm{r} T^{\eta} \right)|\bm{Z}(\bm{r})\rangle.
\end{equation}
In other words,
\begin{equation}
\hat{U}_{\bm{r}+\delta \bm{r}, \bm{r}}=\mathbb{1}+i \hat{{\bm w}} \cdot {\delta \bm{r}}=\mathbb{1}-\frac{i}{4} f_{\mu v \eta} n^{\mu} \hat{T}^{\eta} \partial_{\bm{r}} n^{v}  \cdot \delta \bm{r}.
\end{equation}
To demonstrate this statement, we need to show that the operator field  $\boldsymbol{\mathfrak{n}}(\bm{r})$ is transformed into $\boldsymbol{\mathfrak{n}}(\bm{r}+\delta \bm{r})$
\begin{eqnarray}
\hat{U}_{\bm{r}+\delta \bm{r}, \bm{r}} \boldsymbol{\mathfrak{n}}(\bm{r}) \hat{U}_{\bm{r}, \delta \bm{r}}^{\dagger} &=& \boldsymbol{\mathfrak{n}}(\bm{r}) + \frac{i}{4} f_{\mu \nu \eta} n^{\mu} \partial_{\bm{r}} n^{\nu} \cdot \delta \bm{r} \left[\boldsymbol{\mathfrak{n}}(\bm{r}), \hat{T}^{\eta}\right] =
\boldsymbol{\mathfrak{n}}(\bm{r})+\frac{1}{4} n^{\alpha} n^{\mu}
 f_{\mu \nu \eta}  f_{\alpha \epsilon \eta} \hat{T}^{\epsilon} \partial_{\bm{r}} n^{\nu} \cdot {\delta {\bm r}}
 \nonumber \\
 &=& \boldsymbol{\mathfrak{n}}(\bm{r})+\frac{1}{4} n^{\alpha} n^{\mu} \partial_{\bm{r}} n^{v} \cdot {\delta {\bm r}} \hat{T}^{\epsilon}\left[\frac{8}{3}\left(\delta_{\mu \alpha} \delta_{v \epsilon}-\delta_{\mu \epsilon} \delta_{\alpha v}\right)+4\left(d_{\mu \alpha \eta} d_{v \epsilon \eta}-d_{v \alpha \eta} d_{\mu \epsilon \eta}\right)\right].
\end{eqnarray}
By using the following relationships that can be obtained from the constraint given in Eq.~(11) of the main text:
\begin{eqnarray}
\partial_{\bm{r}} n^{m} n^{m} &=& 0, 
 \\
\partial_{\bm{r}} n^{m} &=& 3 d_{m q p} n^{p} \partial_{\bm{r}} n^{q}, 
 \\
d_{\mu \alpha \eta} n^{\mu} n^{\alpha} &=& \frac{2}{3} n^{\eta},
 \\
\frac{2}{3} n^{\eta} d_{v \epsilon \eta} \partial_{r} n^{\nu} &=& \frac{2}{9} \partial_{r} n^{\epsilon}, 
 \\
d_{v \alpha \eta} \partial_{r} n^{v} n^{\alpha} &=& \frac{1}{3} \partial_{r} n^{\eta}, \\
\frac{1}{3} d_{\epsilon \alpha \eta} n^{\mu} \partial_{r} n^{\eta} &=& \frac{1}{9} \partial_{r} n^{\epsilon},
\end{eqnarray}
we obtain the desired result:
\begin{equation}
\hat{U}_{\bm{r}+\delta \bm{r}, \bm{r}} \boldsymbol{\mathfrak{n}}(\bm{r}) \hat{U}_{\bm{r}+\delta \bm{r}, \bm{r}}^{\dagger}=
\boldsymbol{\mathfrak{n}}(\bm{r})+\partial_{\bm{r}} \boldsymbol{\mathfrak{n}}(\bm{r}) \cdot \delta \bm{r}=\boldsymbol{\mathfrak{n}}(\bm{r}+\delta \bm{r}) .
\end{equation}
Since  $\partial_{\bm{r}}|\bm{Z}\rangle=i \hat{w}_{\delta \bm{r}}|\bm{Z}\rangle$, we have:
\begin{equation}
\begin{aligned}
\partial_{x} \mathcal{A}_{y}-\partial_{y} \mathcal{A}_{x} &=i\left(\partial_{x}\langle \bm{Z}|\right)\left(\partial_{y}|\bm{Z}\rangle-\partial_{y}\langle \bm{Z}|\right)\left(\partial_{x}|\bm{Z}\rangle\right)=\\
&=i\left\langle \bm{Z}\left|\left[\hat{w}_{x}, \hat{w}_{y}\right]\right| \bm{Z}\right\rangle
-\frac{1}{16} f_{\mu \nu \eta} f_{\eta \gamma \epsilon} f_{\alpha \beta \gamma} n^{\mu} n^{\eta} n^{\epsilon}
\partial_x n^{\nu} \partial_y n^{\beta} 
\\
&=\frac{1}{4} f_{\alpha \beta \gamma} \partial_{x} n^{\gamma} \partial_{y} n^{\beta} n^{\alpha} \\
&=-\frac{i}{8} \operatorname{Tr}\left(\boldsymbol{\mathfrak{n}}\left[\partial_{x} \boldsymbol{\mathfrak{n}}, \partial_{y} \boldsymbol{\mathfrak{n}}\right]\right)=-\frac{i}{16} \varepsilon_{j k} \operatorname{Tr}\left(\boldsymbol{\mathfrak{n}}\left[\partial_{j} \boldsymbol{\mathfrak{n}}, \partial_{k} \boldsymbol{\mathfrak{n}}\right]\right),
\end{aligned}
\end{equation}
where we have used the following relationships:
\begin{eqnarray}
n^{\mu} n^{\epsilon} f_{\mu\nu\eta} f_{\gamma \epsilon \eta} &=& \frac{8}{3} (n^{\gamma} n^{\nu} - \frac{4}{3} \delta_{\nu \gamma}) + 4 (n^{\mu} n^{\epsilon} d_{\mu \gamma \eta}  d_{\nu \epsilon \eta} - \frac{2}{3} d_{\nu \gamma \eta} n^{\eta} ) 
, \\
\partial_x n^{\nu} n^{\mu} n^{\epsilon} f_{\mu\nu\eta} f_{\gamma \epsilon \eta} &=& - \frac{32}{9} \partial_{x} n^{\gamma}
+ \frac{4}{3} n^{\mu} d_{\mu \gamma \eta} \partial_x n^{\eta} - \frac{8}{9} \partial_x n^{\gamma}
\nonumber \\
&=&  - \frac{32}{9} \partial_{x} n^{\gamma}
+ \frac{4}{9}  \partial_x n^{\gamma} - \frac{8}{9} \partial_x n^{\gamma} = - 4 \partial_x n^{\gamma}.
\end{eqnarray}
This concludes the demonstration of Eq.~\eqref{eq:prop}. From this result and Eq.~\eqref{eq:skxcharge}, we obtain:
\begin{equation}
C=-\frac{i}{32 \pi} \int \mathrm{d}^{2} x \varepsilon_{j k} \operatorname{Tr}\left(\boldsymbol{\mathfrak{n}}\left[\partial_{j} \boldsymbol{\mathfrak{n}}, \partial_{k} \boldsymbol{\mathfrak{n}}\right]\right)=\frac{1}{2 \pi} \int \mathrm{d} x \mathrm{~d} y\left(\partial_{x} \mathcal{A}_{y}-\partial_{y} \mathcal{A}_{x}\right).
\label{eq:skycharge}
\end{equation}

\subsection{On the lattice}
For lattice systems, the color field is only defined on discrete lattice points. Thus, to compute the skyrmion number of a given
spin configuration we must introduce an interpolation procedure that allows us to to define the spin configuration on any point
of the plane $\mathbb{R}^{2}$. This can be done by connecting color fields $\boldsymbol{\mathfrak{n}}_j$ and  $\boldsymbol{\mathfrak{n}}_k$ on nearest-neighbor sites $j$ 
and $k$ along the geodesic in  CP$^2$. According to this prescription, the contribution to the skyrmion number of a given triangular plaquette $jkl$ of the triangular
lattice is:
\begin{equation}
\rho_{j k l}=-\frac{i}{32 \pi} \int_{\Delta_{j k l}} \mathrm{d}x \mathrm{d}y \varepsilon_{j k} \operatorname{Tr}\left(\boldsymbol{\mathfrak{n}}\left[\partial_{j} \boldsymbol{\mathfrak{n}}, \partial_{k} \boldsymbol{\mathfrak{n}}\right]\right),
\label{eq:skyjkl}
\end{equation}
where $\triangle_{jlk}$ is the triangle formed by the lattice sites $jkl$. Consequently, the total skyrmion number is equal to the sum of this
contribution over all the triangles $jkl$ of the triangular lattice:
\begin{equation}
C=\sum_{\Delta_{j k l}} \rho_{j k l}.
\label{eq:totsky}
\end{equation}
Our next step is to demonstrate that:
\begin{equation}
\rho_{j k l}=\frac{1}{2 \pi}\left(\gamma_{j l}+\gamma_{l k}+\gamma_{k j}\right),
\label{eq:skyBerry}
\end{equation}
where
\begin{equation}
\gamma_{k j}=\arg \left[\left\langle \bm{Z}_{k} \mid \bm{Z}_{j}\right\rangle\right]
\end{equation}
is the Berry connection on the bond $j \to k$ and
\begin{equation}
\gamma_{j l}+\gamma_{l k}+\gamma_{k j}=\oint_{\Delta_{j k l}} \mathcal{A}_{j} d x^{j}
\label{eq:goal}
\end{equation}
is the Berry phase associated with the triangle $jkl$. From Eqs.~\eqref{eq:stokes} and \eqref{eq:skxcharge}, we have
\begin{equation}
\rho_{j k l}=-\frac{i}{32 \pi} \int_{\Delta_{j k l}} \mathrm{d}x \mathrm{d}y \varepsilon_{j k} \operatorname{Tr}\left(\boldsymbol{\mathfrak{n}}\left[\partial_{j} \boldsymbol{\mathfrak{n}}, \partial_{k} \boldsymbol{\mathfrak{n}}\right]\right)=\frac{1}{2 \pi} \oint_{\Delta_{j k l}} \mathcal{A}_{j} d x^{j}.
\end{equation}
Consequently, we just need to demonstrate Eq.~\eqref{eq:skyBerry}.

We first note that, \emph{up to a phase factor} $e^{i \delta \varphi}$, the ``geodesic'' SU(3) spin rotation that connects the coherent states $|\bm{Z}(\bm{r})\rangle$ and
$|\bm{Z}(\bm{r}+\delta \bm{r})\rangle$ is the one given in Eq. (22) and it can be rewritten as
\begin{equation}
e^{i \delta \varphi}|\bm{Z}(\bm{r}+\delta \bm{r})\rangle=\hat{U}_{\bm{r}, \delta \bm{r}}|\bm{Z}(\bm{r})\rangle=\left(\mathbb{1}+\frac{i}{4}\left[\boldsymbol{\mathfrak{n}}(\bm{r}), \partial_{\bm{r}} \boldsymbol{\mathfrak{n}}(\bm{r})\right] \delta \bm{r}\right)|\bm{Z}(\bm{r})\rangle.
\end{equation}
The next observation is that:
\begin{equation}
\bm{\mathfrak{n}}(\bm{r})|\bm{Z}(\bm{r})\rangle=\frac{2}{\sqrt{3}}|\bm{Z}(\bm{r})\rangle
\label{eq:eigen}
\end{equation}
by definition of the coherent state $|\bm{Z}(\bm{r})\rangle$. Consequently, we have
\begin{equation}
e^{i \delta \varphi}=\left\langle \bm{Z}(\bm{r}+\delta \bm{r})\left|\left(\mathbb{1}+\frac{i}{4}\left[\boldsymbol{\mathfrak{n}}(\bm{r}), \partial_{\bm{r}} \boldsymbol{\mathfrak{n}}(\bm{r})\right] \delta \bm{r}\right)\right|\bm{Z}(\bm{r})\right\rangle
\end{equation}
or
\begin{equation}
e^{i \delta \varphi}=\langle \bm{Z}(\bm{r}+\delta \bm{r}) \mid \bm{Z}(\bm{r})\rangle+\left\langle \bm{Z}(\bm{r})\left|\left(\frac{i}{4}\left[\bm{\mathfrak{n}}(\bm{r}), \partial_{\bm{r}} \bm{\mathfrak{n}}(\bm{r})\right] \delta \bm{r}\right)\right|\bm{Z}(\bm{r})\right\rangle=\langle \bm{Z}(\bm{r}+\delta \bm{r}) \mid \bm{Z}(\bm{r})\rangle,
\label{eq:oeverl}
\end{equation}
where we have used that $\left\langle \bm{Z}(\bm{r})\left|\left[\bm{\mathfrak{n}}(\bm{r}), \partial_{r} \bm{\mathfrak{n}}(\bm{r})\right]\right|\bm{Z}(\bm{r})\right\rangle=0$ because of 
Eq.~\eqref{eq:eigen}. This important result shows that, to linear order
in $\delta \bm{r}$, the Berry phase accumulated by the rotation of the coherent state state $|\bm{Z}(\bm{r})\rangle$ along a geodesic of CP$^2$ is equal the the
overlap $\langle \bm{Z}(\bm{r}+\delta \bm{r}) \mid \bm{Z}(\bm{r})\rangle$.

Let us consider now the Berry phase that is obtained when the coherent state $\left|\bm{Z}_{j}\right\rangle$ is rotated into the coherent state $\left|\bm{Z}_{k}\right\rangle$ along
the SU(3) geodesic that connects these two points. After dividing the rotation $\hat{U}_{\bm{r}_{k}, \bm{r}_{j}}$ into a product of $N \to \infty$ small rotations:
\begin{equation}
\hat{U}_{\bm{r}_{k j} / N}=\exp \left\{\frac{i}{4 N}\left[\boldsymbol{\mathfrak{n}}\left(\bm{r}_{j}\right), \boldsymbol{\mathfrak{n}}\left(\bm{r}_{k}\right)\right]\right\},
\end{equation}

\begin{equation}
\begin{aligned}
\gamma_{k j} &=\arg \left[\left\langle \bm{Z}_{k}\left|\hat{U}_{\bm{r}_{k}, \bm{r}_{j}}\right| \bm{Z}_{j}\right\rangle\right]=\arg \left[\left\langle \bm{Z}_{k}\left|\left(\hat{U}_{\bm{r}_{k j} / N}\right)^{N}\right| \bm{Z}_{j}\right\rangle\right] \\
&=\arg \left[\left\langle \bm{Z}_{k}\left|\hat{U}_{\bm{r}_{k j}} / N\right| \bm{Z}\left(\bm{r}_{k}-\bm{r}_{k j} / N\right)\right\rangle\left\langle \bm{Z}\left(\bm{r}_{k}-\bm{r}_{k j} / N\right)\left|\hat{U}_{\bm{r}_{k j} / N} \ldots\right| \bm{Z}\left(\bm{r}_{j}+\bm{r}_{k j} / N\right)\right\rangle\left\langle \bm{Z}\left(\bm{r}_{j}+\bm{r}_{k j} / N\right)\left|\hat{U}_{\bm{r}_{k j}} / N\right| \bm{Z}_{j}\right\rangle\right],
\end{aligned}
\end{equation}
where $\bm{r}_{k j} \equiv \bm{r}_{k}-\bm{r}_{j}$ and the last identity is obtained by inserting expansions of the identity between the unitary operations in
an orthonormal basis that includes the coherent state $\left|\bm{Z}\left(\bm{r}_{j}+n \bm{r}_{k j} / N\right)\right\rangle$ for the identity operator that is inserted on the left of
$\left(\hat{U}_{\bm{r}_{k j} / N}\right)^{n}$ with $1 \leq n \leq N$. After taking the $N \to \infty$ limit and using Eq.~\eqref{eq:oeverl},
\begin{equation}
\lim _{N \rightarrow \infty} \arg \left[\left\langle \bm{Z}\left(\bm{r}_{j}+(n+1) \bm{r}_{k j} / N\right)\left|\hat{U}_{\bm{r}_{k j} / N}\right| \bm{Z}\left(\bm{r}_{j}+n \bm{r}_{k j} / N\right)\right\rangle\right]=\lim _{N \rightarrow \infty} \arg \left[\left\langle \bm{Z}\left(\bm{r}_{j}+(n+1) \bm{r}_{k j} / N\right) \mid \bm{Z}\left(\bm{r}_{j}+n \bm{r}_{k j} / N\right)\right\rangle\right]
\end{equation}
we obtain the desired result:
\begin{equation}
\gamma_{k j}=\arg \left[\left\langle \bm{Z}_{k}\left|\hat{U}_{\bm{r}_{k}, \bm{r}_{j}}\right| \bm{Z}_{j}\right\rangle\right]=\int_{0}^{\left|\bm{r}_{k}-\bm{r}_{j}\right|}\left\langle \bm{Z}(a)\left|\partial_{a}\right|\bm{Z}(a)\right\rangle \mathrm{d} a
\end{equation}
with $|\bm{Z}(a)\rangle=\hat{U}_{\bm{r}_{j}+a \hat{\bm{r}}_{k j}, \bm{r}_{j}}\left|\bm{Z}\left(\bm{r}_{j}\right)\right\rangle$ and $\hat{\bm{r}}_{k j} \equiv \bm{r}_{k}-\bm{r}_{j} /\left|\bm{r}_{k}-\bm{r}_{j}\right|$, which implies Eq.~\eqref{eq:goal}.

\section{Mapping between color field and SU(3) coherent states \label{app:A}}
Eq.~\eqref{eq:skyBerry} is clearly very useful when we are working with coherent states (the $\bm{Z}$-field) instead of working with the color field $\boldsymbol{\mathfrak{n}}(\bm{r})$.
In the latter case, it may be useful to find a coherent state $|\bm{Z}(\bm{r})\rangle$ associated with the field $\boldsymbol{\mathfrak{n}}(\bm{r})$ to keep using the simple formula
provided by Eq.~\eqref{eq:skyBerry} . Note that the coherent state $|\bm{Z}(\bm{r})\rangle$ is defined up to a phase factor (gauge freedom), which does not affect
the value of the Berry phase on a closed loop. Correspondingly, we just need a procedure that allows us to find some state $|\bm{Z}(\bm{r})\rangle$
(a particular gauge choice) for a given $\boldsymbol{\mathfrak{n}}(\bm{r})$.

Eq.~(10) of the main text establishes a mapping $\left|\bm{Z}_{j}\right\rangle \rightarrow \boldsymbol{\mathfrak{n}}(\bm{r})$ between an SU(3) coherent state $|\bm{Z}(\bm{r})\rangle$ and the color field
$\boldsymbol{\mathfrak{n}}(\bm{r})$. Let us find the inverse mapping $ \boldsymbol{\mathfrak{n}}(\bm{r}) \rightarrow  \left | \bm{Z}_{j}\right\rangle$
keeping in mind that a coherent states are defined up to a phase factor (gauge freedom). Given the highest weight state, $| + 1 \rangle$ that satisfies:
\begin{equation}
\hat{T}_{j}^{3}\left|  +1\right\rangle=\frac{2}{\sqrt{3}}\left| +1\right\rangle,
\end{equation}
we can obtain the coherent state $|\bm{Z}(\bm{r})\rangle$ by applying an SU(3) transformation ${\hat U}$ that satisfies
\begin{equation}
\begin{aligned}
\left|\bm{Z}_{j}\right\rangle &=\hat{U}_{j}\left| +1\right\rangle \\
\boldsymbol{\mathfrak{n}}_{j} &=\frac{2}{\sqrt{3}} \hat{U}_{j} \hat{\tilde{T}}_{j}^{3} \hat{U}_{j}^{\dagger}.
\end{aligned}
\end{equation}
This immediately implies that $|\bm{Z}(\bm{r})\rangle$ is the highest-weight eigenstate of the color field $\boldsymbol{\mathfrak{n}}(\bm{r})$ 
\begin{equation}
\boldsymbol{\mathfrak{n}}_{j}\left|\bm{Z}_{j}\right\rangle=\frac{2}{\sqrt{3}}\left|\bm{Z}_{j}\right\rangle,
\end{equation}
and allows us to obtain the coherent state $|\bm{Z}(\bm{r})\rangle$  for given color field $\boldsymbol{\mathfrak{n}}(\bm{r})$. As expected, the normalized eigenstate $|\bm{Z}(\bm{r})\rangle$
is defined up to a multiplicative phase factor.
\bibliographystyle{naturemag}